\documentstyle[12pt]{article}
\global\arraycolsep=2pt %reduces the separation in eqnarrays

\newcommand{\bpipi}{B_d\to\pi^+\pi^-}
\newcommand{\order}{{\cal O}}

\newcommand{\acp}{a_{\mbox{{\tiny CP}}}}
\newcommand{\acpdir}{{\cal A}_{\mbox{{\tiny CP}}}^{\mbox{{\tiny dir}}}}
\newcommand{\acpmi}{{\cal A}_{\mbox{{\tiny
CP}}}^{\mbox{{\tiny mix-ind}}}}
\newcommand{\acc}{A_{\mbox{{\tiny CC}}}}
\newcommand{\apen}{A_{\mbox{{\tiny PEN}}}}
\newcommand{\apenbar}{\overline{A}_{\mbox{{\tiny PEN}}}}

\newcommand{\xif}{\xi_{\pi^+\pi^-}^{(d)}}
\newcommand{\beq}{\begin{equation}}
\newcommand{\eeq}{\end{equation}}
\newcommand{\bea}{\begin{eqnarray}}
\newcommand{\eea}{\end{eqnarray}}

\begin{document}

\thispagestyle{empty}

\rightline{TTP96-49}
\rightline{hep-ph/9610357}
\rightline{October 1996}
\bigskip
\boldmath
\begin{center}
{\Large {\bf Penguin Zoology in $B \to \pi\pi$ and the Extraction}}
\end{center}
\begin{center}
{\Large{\bf of the CKM Angle $\alpha$}}\\
\end{center}
\unboldmath
\smallskip
\begin{center}
{\large{\sc Robert Fleischer}}\footnote{Internet: {\tt 
rf@ttpux1.physik.uni-karlsruhe.de}}\, and \, {\large{\sc Thomas 
Mannel}}\footnote{Internet: {\tt 
tm@ttpux7.physik.uni-karlsruhe.de}}\\
\vspace{0.3cm}
{\sl Institut f\"{u}r Theoretische Teilchenphysik}\\
{\sl Universit\"{a}t Karlsruhe}\\
{\sl D--76128 Karlsruhe, Germany}\\ 
\vspace{0.7cm}
\end{center}
\begin{abstract}
\noindent
We reanalyze the decay $B_d\to\pi^+\pi^-$ without assuming dominance 
of QCD penguins with internal top-quark exchanges. In that case the 
weak phase of the $\bar b \to\bar d$ QCD penguin amplitude is not 
related in a simple way to the CKM angle $\beta$. Nevertheless it is 
still possible to extract the CKM angle $\alpha$ in a transparent and 
almost model-independent way from the time-dependent CP-violating 
asymmetry of the transition $B_d\to\pi^+\pi^-$. Aside from that
CP asymmetry this approach needs as an input only amplitudes of decays 
with branching ratios of order $10^{-5}$ and will thus be well within 
reach at future $B$-factories.
\end{abstract}

\vspace{0.7cm}

\noindent
Among the central goals of the planned dedicated $B$-physics experiments
is the direct measurement of the angles $\alpha$, $\beta$ and $\gamma$
arising in the unitarity triangle (UT) \cite{ut} of the 
Cabibbo--Kobayashi--Maskawa matrix (CKM matrix) \cite{ckm}. Besides the
``gold-plated'' decay $B_d\to J/\psi K_{\mbox{{\scriptsize S}}}$ to determine
$\sin(2\beta)$, the transition $B_d\to\pi^+\pi^-$ plays a very
important role to accomplish this task \cite{cp-revs}. 

It is well-known that the time-dependent CP-violating asymmetry of this
decay defined by 
\bea
\lefteqn{\acp(t)\equiv\frac{\Gamma(B_d^0(t)\to\pi^+\pi^-)-\Gamma(
\overline{B_d^0}(t)\to \pi^+\pi^-)}{\Gamma(B_d^0(t)\to\pi^+\pi^-)+
\Gamma(\overline{B_d^0}(t)\to\pi^+\pi^-)}}\nonumber\\
&&=\acpdir\cos(\Delta M_d\,t)+\acpmi
\sin(\Delta M_d\,t)\label{e1}
\eea
probes the angle $\alpha$ of the UT, where $\Delta M_d>0$ denotes the 
mass splitting of the $B^0_d-\overline{B^0_d}$ mass eigenstates. 
In this expression we have separated direct from mixing-induced CP-violating 
contributions which are characterized by
\beq \label{e2}
\acpdir\equiv\frac{1-\left\vert\xif\right\vert^2}
{1+\left\vert\xif\right\vert^2}, \quad
\acpmi \equiv \frac{2\,\mbox{Im}\,\xif}{1+\left\vert\xif\right\vert^2}.
\eeq
The observable $\xif$ containing 
essentially all the information that is needed to evaluate these quantities
takes the following form:
\beq
\xif
=-e^{-2i\beta} \left[\frac{e^{-2i\gamma}\acc+\apenbar}{\acc+\apen}
\right],\label{e4}
\eeq
where $\acc$ is the contribution of the current-current operators 
including the tree process as well as 
the exchange contribution and $\apen$ ($\apenbar$) are the 
contributions of the penguin type to $B_d^0 \to \pi^+ \pi^-$ 
($\overline{B_d^0} \to \pi^+ \pi^-$). 

If there were no penguin contributions present in $\bpipi$, i.e.\ $\apen=
\apenbar=0$, we would simply have \cite{cp-revs} 
\beq\label{e6}
\acp(t)=-\sin(2\alpha)\sin(\Delta M_d\,t)
\eeq 
and could determine $\alpha$ from that expression. 
However, penguins contribute and disturb the relation 
(\ref{e6}) between $\sin(2\alpha)$ and the CP-violating asymmetry arising
in $\bpipi$. The corresponding hadronic uncertainties have been investigated
by many authors in the previous literature (see e.g.~\cite{pens}). 

One distinguishes between two types of penguin topologies: QCD and
electroweak penguins related to strong and electroweak interactions,
respectively. Consequently the penguin amplitude $\apen$ can be decomposed
as
\beq\label{e7}
\apen=P+P_{\mbox{{\tiny EW}}},
\eeq
where $P$ describes the $\bar b\to\bar d$ QCD penguins and 
$P_{\mbox{{\tiny EW}}}$ denotes the $\bar b\to\bar d$ electroweak penguin 
contributions. Whereas there are certain non-leptonic $B$ decays where 
electroweak penguins play an important role \cite{ewp}, in the transitions 
considered in this letter electroweak penguins enter only in color-suppressed 
form and hence lead to small effects \cite{ewpBpipi,kp}. Therefore we may omit 
the electroweak penguin contributions in the following discussion.  

There are strategies that allow to eliminate the QCD penguin uncertainties in
$\bpipi$ thereby providing a clean extraction of $\alpha$. Probably the best 
known example is the isospin approach based on triangle constructions that 
was proposed by Gronau and London~\cite{gl} and requires in addition to a 
time-dependent measurement of $B_d(t)\to\pi^+\pi^-$ also the branching 
ratios BR$(B^+\to\pi^+\pi^0)$ and BR$(B_d\to\pi^0\pi^0)$. Since the 
branching ratio of $B_d\to\pi^0\pi^0$
is expected to be of $\order(10^{-6})$ or even smaller \cite{kp}, this
approach is -- despite of its attractiveness -- rather difficult from an
experimental point of view. For alternative methods to determine
$\alpha$ see e.g.~\cite{alts}.

In this note we reanalyze the contributions of the QCD penguin type 
to the time-dependent CP asymmtery (\ref{e1}). To the best of our knowledge 
it has been assumed in all previous studies of the hadronic 
uncertainties affecting the determination of $\alpha$ that the QCD penguins 
are dominated by internal top-quark exchanges. Under this assumption 
the $\bar{b} \to \bar{d}$ QCD penguin amplitude $P$ takes the very 
simple form
\begin{equation}
P = e^{-i\beta} e^{i\delta_P} |P|\, ,
\end{equation} 
where $\delta_P$ is a CP-conserving strong phase.

However, as was shown in 
\cite{fbf}, also QCD penguins with up- and charm-quarks running 
as virtual particles in the loops may play an important role and cannot be 
neglected. In particular these penguin topologies contain long-distance 
contributions such as $B^0_d\to\{D^+D^-\}\to\pi^+\pi^-$ (see e.g.\ 
\cite{kamal}) which are very hard to estimate. In the following discussion 
we shall include these contributions into the definition of the 
$\bar b\to\bar d$ QCD penguin amplitudes and hence we relax the condition of 
top-quark dominance. 

Using the notation of \cite{fbf} the contributions of the 
$\bar{b} \to \bar{d}$ QCD penguin type read
\begin{equation}
P = \sum_{q=u,c,t} V_{qd} V_{qb}^\ast P_q \, ,
\end{equation}
where $P_q$ is the amplitude of a QCD penguin process with internal quark
$q$ omitting the CKM factors. Note that $\overline{P}$ 
(the $b \to d$ QCD penguin amplitude) is related to 
$P$ by reversing the signs of the weak phases. 
For the observable $\xi^{(d)}_{\pi^+ \pi^-}$ one 
may write using (\ref{e4})
\begin{equation} \label{amplratio}
\xi^{(d)}_{\pi^+ \pi^-} = 
-\left[\frac{e^{2i\alpha} + e^{-2i\beta} \frac{\overline{P}}{\acc}}
{1 + \frac{P}{\acc}}\right].
\end{equation}
We shall exploit the fact that the penguin contributions $P$
and $\overline{P}$  are expected to be much smaller than the 
current-current amplitude $\acc$ \cite{pen-est,gro}. 
In fact, using the operator basis for the $|\Delta B| = 1$ Hamiltonian 
as given in \cite{heff}, one can obtain a simple estimate for this ratio.
To  this end we shall retain only the top-quark contribution
({\it only} for this estimate!)  and write
\begin{eqnarray}
\overline{P}&=&-\frac{G_{\mbox{{\tiny F}}}}{\sqrt{2}}V_{td}^\ast V_{tb}\sum
\limits_{k=3}^6C_k\langle\pi^+\pi^-|Q_k|\overline{B^0_d}
\rangle\label{estP}\\
\overline{A}_{\mbox{{\tiny CC}}}&=&\frac{G_{\mbox{{\tiny 
F}}}}{\sqrt{2}}V_{ud}^\ast V_{ub}
\sum\limits_{k=1}^2C_k\langle\pi^+\pi^-|Q_k|\overline{B^0_d}
\rangle\label{estA},
\end{eqnarray}
where the $C_k$ are the Wilson coefficients of the current-current 
($k=1,2$) and QCD penguin operators ($k=3,...,6$) taken at the 
scale $\order(m_b)$. The hadronic matrix elements of the four-quark
operators $Q_k$ are evaluated in factorization which implies a Fierz 
re-arrangement for some of the operators. In that way one obtains
\bea 
&& \left|\frac{\overline{P}}{\acc}\right|\approx \left|\frac{P}{\acc}
\right| \nonumber \approx\left|\frac{C_3/N_c + C_4 + 
\frac{2M_\pi^2}{(m_u+m_d)m_b}
[C_5/N_c + C_6]}{C_1/N_c + C_2}\right|
\frac{|V_{td}|}{|V_{ub}|}\nonumber \\ 
&& \qquad\approx0.05\cdot\frac{|V_{td}|}{|V_{ub}|} 
\approx 0.07-0.23, \label{estim}
\eea
where $N_c=3$ is the number of quark-colors and the factor 
$M_\pi^2/[(m_u+m_d)m_b]$ is due to the use of the equations of 
motion for the quark fields. We have used the range for the CKM-factor
$|V_{td}|/|V_{ub}|$ given by Ali and London in Ref.~\cite{al}.

{}From this simple estimate it seems to be justified to expand the 
observable (\ref{amplratio}) in powers of $\overline{P}/\acc$ 
and $P/\acc$ and to keep only the leading term in that expansion, hence
\begin{eqnarray} 
&& \xi^{(d)}_{\pi^+ \pi^-} = -  
\left(e^{2i\alpha} + \frac{e^{-2i\beta}\overline{P}}{\acc}\right)
  \left(1 - \frac{P}{\acc} \right) +
  {\cal O} \left((P/\acc)^2\right)  \nonumber \\ \nonumber  
&& = - e^{2i\alpha} \left[1+\frac{e^{-i\delta_{\mbox{{\tiny CC}}}}}
{|\acc|}\left(e^{i\gamma}\, \overline{P} 
  -  e^{-i\gamma} P \right) \right]+{\cal O}\left( (P/\acc)^2\right),
\end{eqnarray}
where $\delta_{\mbox{{\tiny CC}}}$ denotes the CP-conserving strong phase 
of the amplitude 
\begin{equation}
\acc=e^{i\gamma}e^{i\delta_{\mbox{{\tiny CC}}}}|\acc|. 
\end{equation}
As we will see in a moment, by measuring a ratio of certain 
branching ratios (both $\order(10^{-5})$) it can be tested easily whether 
this expansion is in fact justified. 

In order to discuss the effect of the subdominant penguin 
contributions $P_u$ and $P_c$ we follow \cite{fbf} and write
\begin{equation}
P=\left[ - \frac{1}{R_t} \Delta P + e^{-i\beta} \right] |V_{td}| 
    |P_{tu}| e^{i\delta_{tu}},
\end{equation}
where we have introduced $
P_{q_1 q_2} \equiv |P_{q_1 q_2}| e^{i \delta_{q_1 q_2}}
\equiv P_{q_1} - P_{q_2}$ and 
\beq \label{defs}
R_t \equiv \frac{1}{\lambda} 
            \frac{|V_{td}|}{|V_{cb}|} = \frac{\sin \gamma}{\sin \alpha}
\eeq
with $\lambda\approx0.22$ being the parameter in the Wolfenstein 
parametrization 
\cite{wolf} of the CKM matrix. The quantity $\Delta P$ given by
\begin{equation}
\Delta P \equiv \frac{P_{cu}}{P_{tu}}
\end{equation}
measures the effect of the subdominant penguins $P_u$ and $P_c$. 
Using these relations we obtain
\begin{equation}
e^{i\gamma}\, \overline{P} -  e^{-i\gamma} P =
i\,2\, |V_{td}|\, |P_{tu}| \left(1-\Delta P\right)e^{i\delta_{tu}}\sin\alpha
\end{equation}
{}from which we may express the observable (\ref{amplratio}) as
\beq \label{result}
\xi^{(d)}_{\pi^+ \pi^-} = 
% - e^{2i\alpha} \left[ 1 + 
%i\, 2 \frac{|V_{td}|}{|\acc|}\, e^{-i\delta_{\mbox{{\tiny CC}}}}
%|P_{tu}|\, (1-\Delta P) e^{i\delta_{tu}} \sin \alpha   
%\right] \\
 - e^{2i\alpha} \left[ 1 + i \,2 \lambda R_t e^{i \delta}
\frac{|P'|}{|\acc|} \sin \alpha \right] .
\eeq
Here we have defined an auxiliary quantity 
\begin{equation}\label{Pprime}
P' \equiv |V_{cb}| |P_{tu}|(1-\Delta P) e^{i\delta_{tu}} 
\end{equation}
and $\delta$ is the CP-conserving strong phase of $P'/\acc$. Using (\ref{estP})
and (\ref{estA}), representative values of the Wilson coefficients \cite{heff}
and the factorization assumption to estimate the relevant hadronic matrix 
elements one finds $\delta=0^\circ$. Therefore we expect 
$-90^\circ \le \delta \le 90^\circ$. Furthermore, 
note that $|P'|/|\acc|$ is of order unity since by counting the 
relevant CKM factors and using the estimate (\ref{estim}) we have
\begin{equation}
\frac{|P'|}{|\acc|} \approx \frac{|V_{cb}|}{|V_{td}|} \frac{|P|}{|\acc|}
\approx {\cal O} (1).
\end{equation}

{}From (\ref{result}) one readily obtains the CP-violating asymmetries 
\begin{eqnarray}
&& \acpdir = 
         2 \lambda R_t \frac{|P'|}{|\acc|} 
         \sin \delta \sin \alpha  + {\cal O}\left( (P/\acc)^2\right)
\label{cpasym} \\
&& \acpmi = 
         - \sin 2 \alpha - 2 \lambda R_t \frac{|P'|}{|\acc|}
         \cos \delta\, \cos 2 \alpha\, \sin \alpha + {\cal O} 
\left((P/\acc)^2\right).\label{cpasym1}
\end{eqnarray} 
Hence the two observables $\acpdir$ and $\acpmi$ are 
expressed in terms of three unknown quantities, the ratio $|P'|/|\acc|$, 
the strong phase $\delta$ and UT angle $\alpha$. The presently allowed range
for $R_t$ obtained at 95\% C.L. by constraining the UT using indirect
strategies \cite{al} is $0.68\leq R_t\leq1.55$. Thus $R_t$ is expected
to be of $\order(1)$ within the Standard Model. Strategies for 
extracting the CKM-factor $R_t$ have been discussed e.g.\ in \cite{Rt}.

Note that we have obtained
(\ref{cpasym}), (\ref{cpasym1}) by expanding in $|P|/|\acc|$ 
which is of the same order
as $2 \lambda R_t |P'|/|\acc|$. Keeping only the leading term 
is valid as long as 
the relevant prefactors are of order unity such that the subleading 
terms can savely be neglected. Therefore the approximations made 
allow us to extract $\alpha$ and $\delta$ only in a region in which  
\begin{equation} \label{nottoosmall}
\cos \delta\, \cos 2 \alpha\, \sin \alpha \approx {\cal O} (1),
\end{equation}
where ${\cal O} (1)$ means that this combination of trigonometric functions 
is not as small as $2 \lambda R_t |P'|/|\acc|$. 

Let us note that very similar expressions as those given in 
(\ref{cpasym}), (\ref{cpasym1})
have been derived by Gronau in \cite{gro}. However, we did not assume 
dominance of QCD penguins with internal top-quarks and the physical 
interpretation of the amplitude $P'$ appearing in our formulae is 
quite different from \cite{gro} as we will see in a moment. 
In order to use (\ref{cpasym}), (\ref{cpasym1})  
to determine $\alpha$ one has to 
have some additional input to reduce the number of unknown quantities.
One commonly used input is $SU(3)$ flavor symmetry of strong
interactions \cite{sw} which we shall use in our case to relate the 
magnitude of the auxiliary $\bar b\to\bar d$ amplitude $P'$ (see 
(\ref{Pprime})) to the $\bar b \to \bar s$ QCD penguin amplitude.  

The QCD penguin amplitude involving an $s$-quark instead of a $d$-quark 
takes the form \cite{fbf}
\beq
P^{(s)} =  \sum_{q=u,c,t} V_{qs} V^\ast_{qb}P_q^{(s)} = 
        - \left[1-\Delta P^{(s)}\right] |V_{cb}| \left|P_{tu}^{(s)}\right| 
e^{i\delta_{tu}^{(s)}}
\eeq
and hence exactly corresponds to our auxiliary quantity $(-P')$ in the strict
$SU(3)$ limit, i.e.\ if we do not distinguish between $s$- and $d$-quarks. 
In other words, assuming flavor $SU(3)$, $|P'|$ is simply the magnitude of
the amplitude of a penguin-induced $\bar b \to \bar s$ transition such as
$B^+ \to \pi^+ K^0$. 

Thus one of the unknowns in (\ref{cpasym}), (\ref{cpasym1}) 
{}-- the ratio $|P'|/|\acc|$ --
may be determined with a relatively mild $SU(3)$ input. The magnitude of
the amplitude $|\acc|$ may be obtained from the decay 
$B^+ \to \pi^+ \pi^0$ which receives only color-allowed and 
color-suppressed current-current and negligibly small 
color-suppressed electroweak penguin contributions. Because of the isospin 
symmetry of strong interactions, QCD penguins do not contibute to that 
decay \cite{gl}. If we neglect the color-suppressed current-current and
electroweak penguin contributions to $B^+\to\pi^+\pi^0$ we obtain
\begin{equation}
\left.\frac{|P'|^2}{|\acc|^2}\right\vert_{SU(3)}\approx
\frac{1}{2}\frac{\mbox{BR}(B^+ \to \pi^+ K^0)}{\mbox{BR}(B^+ \to \pi^+ 
\pi^0)},
\end{equation}
where phase space effects have been neglected and the factor 1/2 is due 
to the $\pi^0$-meson. The relevant $B^+$ branching ratios have 
been estimated using the factorization hypothesis to be 
$
\mbox{BR}(B^+ \to \pi^+ K^0) = 9.1\cdot 10^{-6}$ and 
$\mbox{BR}(B^+ \to \pi^+ \pi^0) = 7.8\cdot 10^{-6}
$
\cite{kp} yielding 
\begin{equation}\label{ratio-est}
\left.\frac{|P'|}{|\acc|}\right\vert_{SU(3)} \approx 0.8.
\end{equation}
In the factorization approach $SU(3)$-breaking can be taken into 
account by multiplying with the ratio 
$f_\pi / f_K \approx 0.8$ yielding
\begin{equation} \label{SU3approx}
\frac{|P'|}{|\acc|} \approx \frac{f_\pi}{f_K}\,\sqrt{\frac{1}{2}
\frac{\mbox{BR}(B^+ \to \pi^+ K^0)}{\mbox{BR}(B^+ \to \pi^+ \pi^0)}} 
\end{equation}
which finally allows us a determination of $\alpha$ and $\delta$ from 
the measurement of $\acpdir$ and $\acpmi$ using 
(\ref{cpasym}), (\ref{cpasym1}). 

Let us discuss the uncertainties and limitations of 
this procedure. One limitation is due to the expansion in $|P|/|\acc|$
and has already been given in (\ref{nottoosmall}). The angle 
$\alpha$ is expected to lie within the range $20^\circ \le \alpha 
\le 120^\circ$ \cite{al}
and hence $\sin \alpha$ will not become small. Furthermore, if  
the strong phase $\delta$ is close to the limiting values of 
$\pm 90^\circ$, the extraction of $\alpha$ may be performed by 
using the direct CP-violating asymmetry only since $|\acpdir|$
acquires its maximal value of 
\begin{equation}
|\acpdir | = 2 \lambda R_t \frac{|P'|}{|\acc|} \sin \alpha,
\end{equation}
while our expansion breaks down for $\acpmi$. 
Hence for intermediate values of $\delta$ only the $\cos 2 \alpha$ factor 
in (\ref{nottoosmall}) causes trouble since it vanishes at 
$\alpha = 45^\circ$ and  $\alpha = 135^\circ$. Consequently we have 
to exclude a region $\sigma$ 
around these values in which $|\cos 2 \alpha| \le 2 \lambda R_t |P'|/|\acc|$, 
i.e.\ the subleading terms of the expansion start to become important. From 
this one obtains the exclusion regions 
$45^\circ \pm \sigma$ and $135^\circ \pm \sigma $
with 
\beq
\sigma \approx \frac{180^\circ}{\pi} \lambda R_t \frac{|P'|}{|\acc|} 
\approx 12^\circ R_t \frac{|P'|}{|\acc|}.
\eeq

Flavor $SU(3)$ is generally believed to be good at a level of about
ten to twenty percent. However, as far as $SU(3)$ is concerned, we only 
need it for the relation between branching ratios, while we do not use 
$SU(3)$ for any of the strong phases. Quantitatively we may use 
(\ref{cpasym}), (\ref{cpasym1}) to give an explicit expression 
for the uncertainty 
$\Delta \alpha$ of $\alpha$ that is induced by the uncertainty of 
$|P'|/|\acc|$. We obtain
\begin{equation} \label{error}
|\Delta \alpha | = 
\frac{r \sin^2 \alpha}{2\sqrt{r^2 \sin^2 \alpha - \left(\acpdir
\right)^2}}|\Delta r| \, ,
\end{equation}
where we have used the abbreviation 
\begin{equation} 
r\equiv 2 \lambda R_t \frac{|P'|}{|\acc|} .
\end{equation}

For the following discussion we shall assume that the measurement of the
branching ratios appearing in (\ref{SU3approx}) yields a value of 0.6 for
$|P'|/|\acc|$. This value is in correspondence with (\ref{ratio-est})
if we include factorizable $SU(3)$-breaking through $f_\pi/f_K$.

The ratio $|P'|/|\acc|$ enters our expressions only through $r$. 
In relating the measured branching ratios to $r$ we have to deal with 
uncertainties arising from non-factorizable $SU(3)$-breaking, the neglected 
color-suppressed current-current contributions to $B^+ \to \pi^+ \pi^0$
and from the CKM-factor $R_t$. Assuming an uncertainty in $r$ of 30\%, 
which includes the one from using (\ref{SU3approx}) as well as possible 
uncertainties in $R_t$, is clearly quite conservative. In our examples we 
shall use $r = 0.26$ as the central value corresponding to $|P'|/|\acc| = 
0.6$, $R_t = 1$ and shall not consider any experimental uncertainties.

As the first example let us consider a case where no direct CP violation 
is detected, i.e.\ $\acpdir=0$, and the measurement
yields $\acpmi  = -0.25$.
Ignoring the presence of penguins one would extract from these values 
$\alpha = 83^\circ$; including the penguins along the lines suggested here
one obtains $\alpha = (76 \pm 1)^\circ$, where the uncertainty of one degree
is obtained form (\ref{error}). Note that in this example the QCD penguin
contributions shift the extracted value of $\alpha$ by seven degrees. 

Let us now consider a case with non-vanishing direct CP violation.
Specifically, we assume that the CP asymmetries are measured to be  
$\acpmi = -0.25$ and $\acpdir = +0.1$. In that case one
obtains $\alpha = (76 \pm 1)^\circ$. The non-vanishing direct CP violation 
implies a non-trivial CP-conserving phase $\delta$ which in this 
example turns out to be $\delta = (24 \pm 4)^\circ$. 

In summary, for a 30\% theoretical uncertainty in $r$ and 
for the central value $r=0.26$ the method suggested here is 
expected to allow an extraction of $\alpha$ with an uncertainty of 
$\pm 3^\circ$ if $\alpha$ lies in the allowed regions, i.e.\ not too 
close to $45^\circ$ or $135^\circ$. For values of $\alpha$ far away from 
these singular points one may even have an uncertainty of $\pm 1^\circ$
as has been shown in the above examples.  
Even for larger values of $r$ the method is still quite reliable, 
at least for ``save'' values of $\alpha$.  

Another source of uncertainty is the expansion in 
$|P|/|\acc| = {\cal O}(r)$. To give some idea of the effect of the 
higher order terms in $|P|/|\acc|$ we shall use our first example. 
Here the linear terms in $|P|/|\acc|$ induced a shift of seven degrees 
which amounts to roughly ten percent. Therefore we expect the corrections 
from second order terms to be of the order of a percent, i.e.\ in absolute 
values of about one degree. 

Since the approach presented in this letter
requires neither difficult measurements of very small branching
ratios nor complicated geometrical constructions it may turn out to
be very useful for the early days of the $B$-factory era beginning at
the end of this millennium.

\end{document}